\documentclass[prl,showpacs,showkeys,floatfix,nofootinbib,reprint,twocolumn]{revtex4-1}

\usepackage{graphicx}
\usepackage{amsmath}
\usepackage{amsfonts}
\usepackage{bm}
\usepackage{bbm}
\usepackage{url}
\usepackage{amssymb}
\usepackage{mathrsfs}
\usepackage{dsfont}
\usepackage{subfigure}
\usepackage{paralist}

\usepackage{stmaryrd}

\newcommand{\real}{\mathbb{R}}

\newcommand{\q}{\mathsf{q}}
\newcommand{\Q}{\mathsf{Q}}

\newcommand{\qdot}{\dot{\mathsf{q}}}
\newcommand{\gforce}{\frac{\delta L}{\delta\q}}
\newcommand{\ginertia}{\frac{\delta T}{\delta\q}}
\newcommand{\rforce}{\frac{\partial R}{\partial\qdot}}

\begin{document}
\title{Comment on ``Classical Mechanics of Nonconservative Systems''}
\author{Epifanio G. \surname{Virga}}
\email[e-mail: ]{eg.virga@unipv.it}
\affiliation{Dipartimento di Matematica, Universit\`a di Pavia, Via Ferrata 5, I-27100 Pavia, Italy}

\date{\today}

\begin{abstract}
A Comment on the Letter by C. R. Galley, Phys. Rev. Lett. 110, 174301 (2013).
\end{abstract}



\maketitle

Dissipative systems are ubiquitous in physics but their description in general mathematical terms is still a debated issue. Galley~\cite{galley:classical} proposed a general approach to the motion of a discrete dissipative system based on a modified Hamilton principle, which remedies the time-reversibility of the Lagrange equations, which express the stationarity of the classical action. The major conceptual drift for developing this method was the supposed inability of the classical Rayleigh equations to account for resistive forces more general than linear functions in the velocities. It is my intention to show that a (slight) extension of Rayleigh classical formalism is able to encompass general dissipative potentials, also amenable to a variational formulation.

The notation employed here for a discrete dynamical system is standard. The generalized coordinates form the vector $\q=(q_1,\dots,q_m)\in\real^m$; correspondingly, $\qdot=(\dot{q}_1,\dots,\dot{q}_m)$ is the vector of generalized velocities. $T(\q,\qdot)$ is the kinetic energy, taken to be a quadratic positive definite form in $\qdot$; $V(\q)$ is the potential energy of all conservative generalized forces $\Q$, so that $\Q=-\frac{\partial V}{\partial\q}$, and $L(\q,\qdot):=T-V$ is the Lagrangian of the system, which enjoys the usual smoothness assumptions. Letting $H(\q,\qdot):=T+V$ be the \emph{total energy}, it is a plain consequence of $T$ being quadratic in $\qdot$ that (see also \cite[p.\,125]{sonnet:dissipative}) $\dot{H}+W=0$, where
\begin{equation}\label{eq:W_definition}
W:=\qdot\cdot\gforce=\sum_{j=1}^m\dot{q}_j\left[\frac{\partial L}{\partial q_j}-\frac{d}{dt}\left(\frac{\partial L}{\partial\dot{q}_j}\right)\right]=\qdot\cdot\left(\Q+\ginertia\right)
\end{equation}
is the \emph{total mechanical power} expended by active ($\Q$) and inertial ($\ginertia$) forces. Were these the only forces at work, by the Lagrange equations, $\gforce$ would vanish identically along any motion, and so would $W$, implying that the total energy is conserved.

We assume that nonconservative generalized forces can be expressed as $-\rforce$, where $R(\q,\qdot)$ is a dissipation \emph{potential}, \emph{not} necessarily \emph{quadratic} in $\qdot$. Generalized Lagrange-Rayleigh equations can then be written in the traditional textbook form, $\gforce-\rforce=0$, which combined with \eqref{eq:W_definition} transform the balance of energy into
\begin{equation}\label{eq:D_definition}
\dot{H}=-\qdot\cdot\rforce=:-D,
\end{equation}
where $D(\q,\qdot)$ is to be interpreted as the \emph{dissipation} in the system. Here $D$ is the \emph{only} constitutive function for the nonconservative forces: it is \emph{positive} semidefinite in $\qdot$, but \emph{not} necessarily \emph{quadratic}. The dissipation potential $R$ is determined in terms of $D$ as a solution to the partial differential equation in \eqref{eq:D_definition}$_2$. By applying the method of characteristics to this equation \cite[Ch.\,II]{courant:methods}, we easily arrive at the following explicit representation for $R$ (to within an arbitrary constant):
\begin{equation}\label{eq:R_representation}
R(\q,\qdot)=\left.\int D(\q,e^s\qdot)ds\right|_{s=0}.
\end{equation}

A number of consequences can be drawn from \eqref{eq:R_representation}. First, if $D$ is a homogeneous function of degree $n$, then $R=\frac1nD$, which for $n=2$ reproduces the classical relationship between Rayleigh potential and dissipation. Second, if more generally $D=\sum_{n=1}^ND_n$, with $D_n$ homogeneous of degree $n$ in $\qdot$, then by \eqref{eq:R_representation} $R=\sum_{n=1}^N\frac1nD_n$. In particular, letting $D=A(\q)|\qdot|^3$, with $A(\q)\geqq0$, we recover the quadratic dependence on the velocity typical of the resistive drag opposed to bodies by air flows at high Reynolds numbers \cite[Sec.\,5.11]{batchelor:introduction}, an example similarly encompassed by the method proposed in \cite{galley:classical}.

The generalized Rayleigh-Lagrange equations also admit a variational formulation, which following \cite[p.\,122]{sonnet:dissipative} we call the principle of \emph{reduced dissipation}. It states that for a discrete system with total mechanical power $W$ as in \eqref{eq:W_definition} and dissipation potential $R$ that obeys \eqref{eq:D_definition}$_2$ the \emph{true} velocity $\qdot$ traversing a given configuration $\q$ is such that the \emph{reduced} dissipation potential $\widetilde{R}:=R-W$ is stationary with respect to all \emph{virtual} variations $\delta\qdot$ once the generalized forces $\gforce$ are held fixed.\footnote{For a quadratic dissipation potential $R$, this principle was essentially formulated by Biot~\cite{biot:variational} and a justification was also provided in \cite{sonnet:dynamics}. It can be regarded as the natural extension to dissipative systems of the classical d'Alembert principle.}

In conclusion, I have shown that a generalized dissipation potential derived from a (non necessarily quadratic) dissipation function can be set as the basis of the Rayleigh-Lagrange dynamics of a discrete system, which may thus remain a viable alternative to the approach proposed in \cite{galley:classical}. The method suggested in this Comment is likely to be extended to continuum systems along the lines followed in the monograph \cite{sonnet:dissipative} for the classical quadratic case.

\begin{acknowledgements}
I am indebted to James A. Hanna, who in the course of a stimulating, ongoing correspondence drew my attention to \cite{galley:classical}.
\end{acknowledgements}


\begin{thebibliography}{6}
\expandafter\ifx\csname natexlab\endcsname\relax\def\natexlab#1{#1}\fi
\expandafter\ifx\csname bibnamefont\endcsname\relax
  \def\bibnamefont#1{#1}\fi
\expandafter\ifx\csname bibfnamefont\endcsname\relax
  \def\bibfnamefont#1{#1}\fi
\expandafter\ifx\csname citenamefont\endcsname\relax
  \def\citenamefont#1{#1}\fi
\expandafter\ifx\csname url\endcsname\relax
  \def\url#1{\texttt{#1}}\fi
\expandafter\ifx\csname urlprefix\endcsname\relax\def\urlprefix{URL }\fi
\providecommand{\bibinfo}[2]{#2}
\providecommand{\eprint}[2][]{\url{#2}}

\bibitem[{\citenamefont{Galley}(2013)}]{galley:classical}
\bibinfo{author}{\bibfnamefont{C.~R.} \bibnamefont{Galley}},
  \bibinfo{journal}{Phys. Rev. Lett.} \textbf{\bibinfo{volume}{110}},
  \bibinfo{pages}{174301} (\bibinfo{year}{2013}).

\bibitem[{\citenamefont{Sonnet and Virga}(2012)}]{sonnet:dissipative}
\bibinfo{author}{\bibfnamefont{A.~M.} \bibnamefont{Sonnet}} \bibnamefont{and}
  \bibinfo{author}{\bibfnamefont{E.~G.} \bibnamefont{Virga}},
  \emph{\bibinfo{title}{Dissipative Ordered Fluids: Theories for Liquid
  Crystals}} (\bibinfo{publisher}{Springer}, \bibinfo{address}{New York},
  \bibinfo{year}{2012}).

\bibitem[{\citenamefont{Courant and Hilbert}(1962)}]{courant:methods}
\bibinfo{author}{\bibfnamefont{R.}~\bibnamefont{Courant}} \bibnamefont{and}
  \bibinfo{author}{\bibfnamefont{D.}~\bibnamefont{Hilbert}},
  \emph{\bibinfo{title}{Methods of Mathematical Physics}}
  (\bibinfo{publisher}{Wiley}, \bibinfo{address}{New York},
  \bibinfo{year}{1962}), \bibinfo{note}{{V}ol. II. Partial Differential
  Equations}.

\bibitem[{\citenamefont{Batchelor}(1967)}]{batchelor:introduction}
\bibinfo{author}{\bibfnamefont{G.~K.} \bibnamefont{Batchelor}},
  \emph{\bibinfo{title}{An Introduction to Fluid Dynamics}}
  (\bibinfo{publisher}{Cambridge University Press},
  \bibinfo{address}{Cambridge}, \bibinfo{year}{1967}).

\bibitem[{\citenamefont{Biot}(1955)}]{biot:variational}
\bibinfo{author}{\bibfnamefont{M.~A.} \bibnamefont{Biot}},
  \bibinfo{journal}{Phys. Rev.} \textbf{\bibinfo{volume}{97}},
  \bibinfo{pages}{1463} (\bibinfo{year}{1955}).

\bibitem[{\citenamefont{Sonnet and Virga}(2001)}]{sonnet:dynamics}
\bibinfo{author}{\bibfnamefont{A.~M.} \bibnamefont{Sonnet}} \bibnamefont{and}
  \bibinfo{author}{\bibfnamefont{E.~G.} \bibnamefont{Virga}},
  \bibinfo{journal}{Phys. Rev. E} \textbf{\bibinfo{volume}{64}},
  \bibinfo{pages}{031705} (\bibinfo{year}{2001}).

\end{thebibliography}

\end{document}